\documentclass{elsart3}

\usepackage{graphics}
\usepackage{epsfig}

\usepackage{amssymb}

\begin{document}

\begin{frontmatter}

\title{Pinning Induced Fluctuations on Driven Vortices}

\author{Alejandro B. Kolton}

\address{Universit\'e de Gen\`eve, DPMC, 24 Quai 
Ernest Ansermet, CH-1211 Gen\`eve 4, Switzerland}

\begin{abstract}
We use a simple model to study  
the long time fluctuations induced by random pinning 
on the motion of driven 
non--interacting vortices. We find that  
vortex motion seen from the co--moving frame is 
diffusive and anisotropic, with velocity 
dependent diffusion constants. Longitudinal and transverse 
diffusion constants cross at a characteristic 
velocity where diffusion  
is isotropic. The diffusion front is  
elongated in the direction of the drive 
at low velocities and elongated in the 
transverse direction at large 
velocities. We find that the mobility in the 
driven direction is always larger than the 
transverse mobility, and becomes isotropic 
only in the large velocity limit. 
\end{abstract}

\begin{keyword}
Random Pinning \sep Diffusion \sep Mobility
\PACS 74.25.Qt
\end{keyword}
\end{frontmatter}
\section{INTRODUCTION}
Understanding the non--equilibrium statistical properties 
of driven particles in disordered media is a challenging 
question relevant to many experimental situations. 
A prominent example are the moving phases 
of driven vortex lattices in superconductors 
\cite{GLD}. A key feature of these systems  
is that the disorder induces anisotropic response and 
fluctuations which are strongly 
controlled by the velocity \cite{GLD,KV,kolton}.

In spite of its relevance to understand 
situations of incoherent or plastic vortex flow the 
simple case of an isolated vortex driven in a 
$d$-dimensional random potential with $d>1$ 
has been tackled analitically only by 
perturbation theory \cite{KV}, valid at 
high velocities, or by mean field 
theory \cite{Horner}, valid for $d \gg 1$. In 
this paper we propose a simple model 
whose long time behaviour can be computed analitically 
at $d=2$ for any finite velocity. 

\section{MODEL}
Let us consider the equation of motion, in two dimensions, 
of a driven isolated vortex at zero temperature,
\begin{equation}
\eta {\bf v}(t) = {\bf F} + \sum_i {\bf f}_p({\bf r}-{\bf r}_{i})
\label{eq:ecmov1}
\end{equation}
where ${\bf v}=d{\bf r}/dt$ is the instantaneous velocity of the 
vortex located at ${\bf r(t)}$, ${\bf f}$ is the driving force and 
$\eta$ the friction coefficient. 
We model the disorder as a random arrangement of hard disks with 
center ${\bf r}_i$ and radius $\xi$. Outside the disks the vortex 
has a free motion and inside it feels a pinning force,  
\begin{equation}
{\bf f}_p({\bf r}) = -\frac{A_p}{\xi^2} {\bf r} \Theta(1-r^2/\xi^2) 
\end{equation}
where $A_p$ is the amplitude, and $\Theta$ the step function. 
This disorder models a diluted distribution of 
pinning centers separated at a distance $d > 2\xi$. In 
the following we use adimensionalized variables: $\xi$ is 
the length unit, $A_p$ the energy unit, 
and $A_p/\xi^2 \eta$ the time unity.

Above the depinning transition the motion 
of the vortex consists in straight segments of 
free motion interrupted by the collisions with 
the different pinning centers. At each collision  
the vortex is delayed and deflected with respect to the 
free motion. The equation describing the motion 
of the vortex inside the trap centered at ${\bf r}_i=0$ 
is, 
\begin{equation}
\frac{d \bf r}{dt} = - {\bf r}+{\bf F}
\label{eq:eqcol}
\end{equation}
The collision starts with the vortex 
at some initial position ${\bf r}(0)={\bf r}_0$ on 
the border of the trap, $r^2_0=1$. The 
solution of Eq.(\ref{eq:eqcol}) is 
\begin{equation}
{\bf r}(t)=({\bf r}_0-{\bf F})e^{-t}+{\bf F}
\label{eq:soltrap}
\end{equation}
After a time interval $\delta t$ the vortex will exit from the trap, 
therefore $r^2(\delta t)=1,\,\delta t>0$. Using 
this condition in equation \ref{eq:soltrap} we obtain 
the following expression for $\delta t$,
\begin{eqnarray}
e^{-\delta t} =  
\frac{{\bf f}_p^0.{\bf F} - \sqrt{[{\bf f}_p^0.{\bf F}]^2-
(F^2-1)({\bf f}_p^0)^2}}{({\bf f}_p^0)^2} 
\label{eq:expdt}
\end{eqnarray}
where ${\bf f}_p^0={\bf f}_p({\bf r}_0)$. 
The displacement $\delta {\bf r}$ induced by the collision is 
then given by, 
\begin{equation}
\delta {\bf r} \equiv {\bf r}(\delta t)-{\bf r}_0 = 
({\bf r}_0-{\bf F})(e^{-\delta t}-1)
\label{eq:dxi}
\end{equation}
Due to the random distribution of pinning centers, 
the motion can be considered as a random walk in the long time 
limit. The fluctuations of ${\bf r}(t)$ are thus induced
by the uncorrelated sequence of collisions.
Assuming identical pinning 
centers, the randomness comes exclusively from the 
random initial conditions of each collision, 
described by Eq.(\ref{eq:soltrap}).  

In a long time interval $t$ the vortex collides with a 
large number $N_c(t)$ of pinning centers. Since the core 
covers an area $\sim V t \xi$, where $V$ is 
the mean velocity, we get, 
\begin{equation}
N_c(t)\approx n_p \xi V t. 
\label{eq:nrocollisions}
\end{equation}
By symmetry, the long time displacement is along the 
direction of ${\bf F}$, 
$\Delta {\bf r}(t)=\Delta r_{\parallel}(t) \hat{F}$, and 
\begin{equation}
\Delta r_{\parallel}(t) \equiv V t \approx N_c(t)\langle \delta r_{\parallel} \rangle 
+ F(t- N_c(t)\langle \delta t \rangle )
\label{eq:Dxi}
\end{equation}
where $\langle ... \rangle$ denotes an average over 
a random distribution of initial conditions ${\bf r}_0$ 
in equations \ref{eq:expdt} and \ref{eq:dxi}. From Eq. 
(\ref{eq:Dxi}) we get the mean velocity, 
\begin{eqnarray}
V &=& \frac{F}{1-n_p \xi[\langle \delta r_{\parallel}\rangle - F \langle \delta t\rangle]}
\label{eq:V}
\end{eqnarray}

We can now define longitudinal $D_{\parallel}$ and transverse $D_{\perp}$ 
diffusion constants,
\begin{eqnarray}
D_{\parallel} &\equiv& \langle [\Delta r_{\parallel}(t)-Vt]^2 \rangle /t  \\
D_{\perp} &\equiv& \langle \Delta r_{\perp}^2(t) \rangle /t 
=\frac{N_c(t)}{t}\langle \delta r_{\perp}^2 \rangle 
\label{eq:difutrans}
\end{eqnarray}
To calculate the longitudinal diffusion constant in terms of the 
single collision displacement we 
use that $d- V (F \delta t + d - \delta r_{\parallel})/F$ is the random 
longitudinal displacement with respect to the average longitudinal 
motion in a single collision, 
with $d = n_p^{-1/2}$ the longitudinal mean distance between 
pinning centers of two given consecutive collisions. We thus get, 
\begin{equation}
D_{\parallel}=  \frac{N_c(t)}{t} 
\langle \bigl[ d- \frac{V}{F}(F \delta t + d - \delta r_{\parallel}) \bigr]^2 \rangle
\label{eq:difulong}
\end{equation}
We also define the longitudinal mobility as 
$\mu_{\parallel} \equiv [\frac{d V}{d F}]$.  
In terms of single collision quantities we get,
\begin{equation}
\mu_{\parallel} = 
\frac{V}{F}\biggl\{ 1 - V n_p \xi \biggl( 
\langle \delta t \rangle + F \frac{d \langle \delta t \rangle}{dF} 
- \frac{d \langle \delta r_{\parallel} \rangle}{dF} \biggr)\biggr\} 
\label{eq:mupara}
\end{equation}
To define the transverse mobility we need to introduce 
a small perturbative force $f_{\perp}$. The velocity induced 
by this force is,
\begin{eqnarray}
v_{\perp} &=& V n_p \xi [\langle \delta r_{\perp} \rangle - 
f_{\perp} \langle \delta t\rangle] + f_{\perp}
\label{eq:Vtrans}
\end{eqnarray}
and thus we can define the transverse mobility 
$\mu_{\perp} \equiv [\frac{d v_{\perp}}{d f_{\perp} }
]_{f_{\perp}\rightarrow 0}$.
In terms of single collision quantities we get,
\begin{equation}
\mu_{\perp} = 1 + V n_p \xi \biggl( 
\frac{d \langle \delta r_{\perp} \rangle}{df_{\perp}} 
- \langle \delta t \rangle 
\biggr)_{f_{\perp}\rightarrow 0} 
\label{eq:muperp} 
\end{equation}
Finally we can define effective 
temperatures using generalized Einstein relations, 
\begin{eqnarray}
T_{\tt eff}^{\perp}&=&D_{\perp}/2 \mu_{\perp}\\
T_{\tt eff}^{\parallel}&=&D_{\parallel}/2 \mu_{\parallel}
\end{eqnarray}
In order to calculate the transport properties defined above 
($V$, $D_{\perp}$, $D_{\parallel}$, $\mu_{\perp}$, 
$\mu_{\parallel}$, $T_{\tt eff}^{\perp}$, and 
$T_{\tt eff}^{\parallel}$) we need to calculate 
the first moments of the distribution of 
$\delta t$, $d (\delta t) /d F$,  $\delta {\bf r}$, 
$\delta r_{\parallel} \delta t$, $d (\delta x_{\perp}) /d f_{\perp}$, 
$d (\delta x_{\parallel}) /d F$ by 
performing simple integrals.
\begin{figure} 
\centerline{\includegraphics[height=8.0cm]{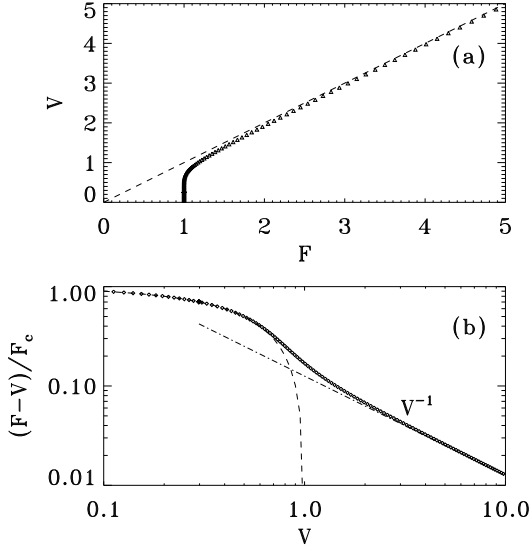}}
\caption{
(a) VF characteristics of the driven particle. The dashed--line 
is the free flow solution. (b) Correction to the free flow velocity. 
Lines show the asymptotic forms at small and large velocities.
}
\label{fig:VF}
\end{figure}
\section{RESULTS}
In Fig. \ref{fig:VF}(a) we show the Velocity - Force characteristics (VF) 
of our model, calculated from Eq.(\ref{eq:V}). The critical 
depinning force is 
$F_c=1$. At zero temperature, for $F<F_c$, 
the particle is trapped after a transient, 
and thus $V=0$. At low velocities, $F \rightarrow F_c^+$, 
the VF curve is strongly nonlinear with 
$V \sim [n_p \log ( (F-F_c)/2F_c )]^{-1}$  \cite{kolton_unpublished} 
and at large velocities free flux flow is approached with corrections 
that scale as $F-V \sim V^{-1}$, as shown in Fig.\ref{fig:VF}(b). 
\begin{figure}
\centerline{\includegraphics[height=8.0cm]{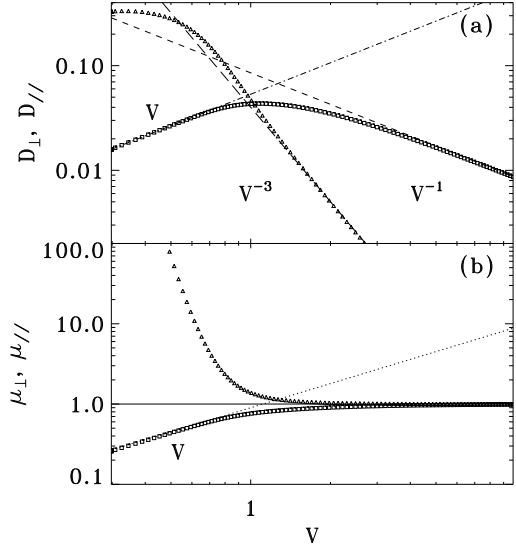}}
\caption{
(a) Diffusion constants in the longitudinal ($\triangle$) 
and transverse ($\Box$) directions.(b) Mobility in the 
transverse ($\Box$) and
longitudinal ($\triangle$) directions. Lines 
show the small and large velocity asymptotic forms in both figures. }
\label{fig:difyres}
\end{figure}
In Fig. \ref{fig:difyres}(a) we show the longitudinal and transverse 
diffusion constants. $D_{\parallel}$ and $D_{\perp}$ are both 
non--monotonous functions of the
velocity $V$. At small velocity diffusion constants grow 
linearly with $V$ while at large
velocity $D_{\parallel} \sim V^{-3}$ and $D_{\perp} \sim V^{-1}$. 
Let us note that $D_{\parallel}$ and $D_{\perp}$ cross at a characteristic 
velocity $V_{\circ}$. This crossing means that 
the long time diffusion front changes aspect ratio at $V_{\circ}$. For 
$V<V_{\circ}$ the diffusion front is elongated in the driven direction, 
while for $V<V_{\circ}$ is elongated in the transverse direction. At 
$V_{\circ}$ diffusion is isotropic. Interestingly, 
the same behavior is observed in 
numerical simulations of interacting vortices in two 
dimensions \cite{kolton}.

In Fig. \ref{fig:difyres}(b) we show the longitudinal and
transverse mobilities. We observe that both are velocity dependent and 
approach the free flux response, $\mu \sim 1$, at large velocities. Since 
the $\mu_{\parallel}$ is the differential resistance, $dV/dF$, 
the divergence observed at small 
velocity is a signature of the depinning 
transition. $\mu_{\parallel}$ near depinning is high 
since any small force can 
reduce strongly the waiting time $\delta t \sim 1/V$ inside the trap. 
On the contrary, $\mu_{\perp} \sim V$ at small 
velocity. A small transverse force has a small effect in the trap 
time $\delta t$ compared with the linear $V$ dependence 
(Eq.(\ref{eq:nrocollisions})) of the number of collisions 
per unit time.	 

\begin{figure}
\centerline{\includegraphics*[height=8.0cm]{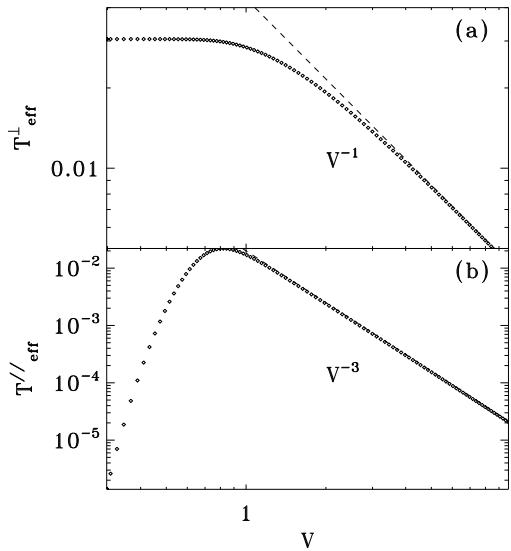}}
\caption{Effective temperatures in the transverse (a) and longitudinal (b) 
directions. Lines indicate the asymptotic forms at large 
velocity.}
\label{fig:teff}
\end{figure}

In Fig. \ref{fig:teff} we show the effective temperatures in the 
longitudinal and transverse directions. Since at large velocities the 
mobilities saturate, the 
velocity dependence of the effective temperatures are dominated 
by the diffusion constants, and thus 
$T^{\perp}_{\tt eff}\sim 1/V$ and $T^{\parallel}_{\tt eff}\sim 1/V^3$. 
At low velocities the effective temperatures 
are instead strongly determined by the velocity dependence of the 
mobilities. $T^{\parallel}_{\tt eff}$ reaches a maximum at a 
intermediate velocity and decreases quickly to zero 
with decreasing $V$. On the other hand, $T^{\perp}_{\tt eff}$ 
saturates at a finite value as the velocity vanishes, due to the linear 
dependence with velocity of both the diffusion constant and the 
mobility.

It is worth noting that only at large velocities and in the transverse
direction, the effective temperature is 
found to be identical, except from numerical factors, 
to the shaking temperature 
\cite{KV}. This confirms that the shaking temperature 
is equivalent to an effective temperature defined from 
a generalized fluctuation relation only in the limit of large 
velocity of non--interacting vortices \cite{kolton}. It is important 
to point out here that the thermodynamic 
nature of the effective temperatures of this model is 
still unclear, since the system is non--interacting and 
strongly driven \cite{CKP}.
\section{CONCLUSIONS}
We have studied the pinning--induced anisotropic 
diffusion of driven non--interacting vortices. 
We find that the diffusion front 
is elongated in the direction of the driving force 
at low velocities and in the transverse direction at 
large velocities. This implies the existence of 
isotropic diffusion at a characteristic velocity 
$V_{\circ}$. The analysis of the anisotropic low frequency 
voltage noise in superconductors could be a 
possible experimental probe of this result, since 
diffusion constants and velocity fluctuations 
are related by generalized Green-Kubo relations 
\cite{kolton}. 

Even if the depinning 
transition does depend on the 
peculiarities of our model, we find that the main 
features of the long time fluctuations we obtain at intermediate 
and large velocities are in agreement with perturbation 
theory predictions \cite{KV} and with numerical 
simulations of non interacting vortices for 
the non--simplified model \cite{kolton_unpublished}. 
Furthermore, it is easy to show that 
our model can be solved at any dimension $d>1$ 
and generalized to more complicated short-range 
random potentials \cite{kolton_unpublished}. 
 
We acknowledge discussions with A. Rosso,
A. Iucci, T. Giamarchi, and  D. Dom\'{\i}nguez. 
This work was supported in 
part by the Swiss National Fund under Division II.

\end{document}